# DESIGN AND IMPLEMENTATION OF THE END SYSTEM TO INTERMEDIATE SYSTEM (ES-IS) ROUTING INFORMATION EXCHANGE PROTOCOL AS A LOADABLE KERNEL MODULE IN LINUX KERNEL 2.6

Stella Maria[1], Maulahikmah Galinium[1], Husni Fahmi[2],
Haret Faidah[2], James Purnama[3], Charles Lim[3], Harya Damar[3]

**Abstract**
*This paper presents a partial implementation of the ES-IS Routing Information Exchange Protocol packet processing in Linux Kernel 2.6, which is for use in conjunction with the Connectionless Network Protocol (CLNP) in Aeronautical Telecommunication Network (ATN). First, we show the data structures involved in the protocol operation. Second, we describe the map of the packet processing whose design has been developed in the research. Third, we explain how the protocol is implemented as a loadable kernel module. Finally, we conclude the implementation result based on performed tests.*
**Keywords**: *CNS/ATM, ATN, CLNP, ES-IS Routing, ICAO*

## 1. Introduction

The increasing demand of air travel requires the implementation of a new technology to manage the air space for more optimum utilization while improving safety level. Therefore, the International Civil Aviation Organization (ICAO) tasked its member countries, including Indonesia, to apply the standardized Communications, Navigation, Surveillance, and Air Traffic Management (CNS/ATM) systems [4]. Acting upon this mandate, the Agency for the Assessment and Application of Technology (BPPT) initiated the research to help satisfy the needs of the Directorate General of Civil Aviation, and PT Angkasa Pura I and II to implement those systems in Indonesia [10].

BPPT began the project with the development of the ATN [9]. This infrastructure enhances the communications aspect with ground-ground and air-ground links for digital data transfer between aircraft and civil air traffic control facilities. In ATN, data is conveyed by the CLNP [5], which lies on the network layer together with several routing information exchange protocols.

As a part of the project, we have built the prototype of the ES-IS Routing Information Exchange Protocol as a loadable kernel module in Linux Kernel 2.6. The implemented protocol has been supplied with the capabilities to provide basic configuration information that allows ESs and ISs directly connected through the same subnetwork to discover each other's presence. However, the functions for the distribution of route redirection information are not yet built. Furthermore, there has been no test for reliability and performance measurement conducted as of now.

[1] Swiss German University, German Center, Tangerang, Indonesia, {stella.maria, maulahikmah.galinium}@student.sgu.ac.id
[2] The Agency for the Assessment and Application of Technology/Badan Pengkajian dan Penerapan Teknologi (BPPT), Jakarta, Indonesia, {fahmi, haret}@inn.bppt.go.id
[3] Department of Information Technology, Swiss German University, German Center, Tangerang, Indonesia, {james.purnama, charles.lim, harya.widiputra}@sgu.ac.id

## 2. Related Work

The routing protocols in CLNP are different from those in IPv4. As the intra domain routing protocols, CLNP uses ES-IS and IS-IS Routing Information Exchange Protocols. On the other hand, IPv4 uses Routing Information Protocol (RIP) and Open Shortest Path First (OSPF) [2]. RIP is required to help routers dynamically adapt to changes of network connections by communicating information of which networks each router can reach and their distances. It uses distance-vector algorithm, which employs the number of hops as a routing metric [8]. The OSPF uses Dijkstra Algorithm to find the shortest path. Still, the ES-IS Routing Information Exchange Protocol is developed by following the modular design of IPv4 in existing Linux kernel [8].

The CLNP and ES-IS Routing Information Exchange Protocol have been once implemented in NetBSD. The complete source code is provided in NetBSD CVS Repository [7], and the source for the ES-IS Routing Information Exchange Protocol can be found is *esis.c* file. In this file, the *esis_config* method performs report configuration function. Then, the constructed ESH or ISH PDU is transmitted to device by the *esis_shoutput* method. Any incoming packet is processed by the *esis_input* method. When the received packet is an ESH PDU, it is handled further by the *esis_eshinput* method. On the other hand, *esis_ishinput* method processes received ISH PDU. Finally, *esis_insert_addr* method performs the record configuration function. Compared to our implementation, the one in NetBSD already includes functions for route redirection information.

## 3. General Data Structures

### 3.1. PDU Structures

According to ISO 9542 [6], each type of PDU in the protocol must contain fixed and address parts. Only RA PDU does not have address part. The fixed part carries general information to manage the PDU itself. Apart from that, the content of address part depends on what routing information to be sent. Additionally, a PDU may or may not have options part. The PDUs are as listed below:
a.  ESH (End System Hello) PDU

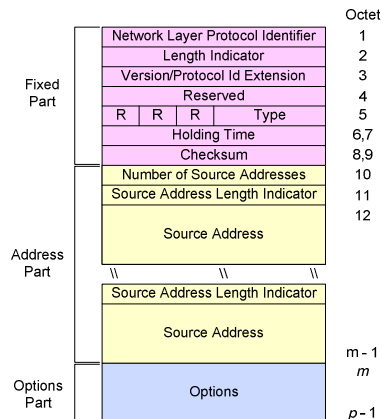

**Figure 1. ESH PDU [6]**

Figure 1 illustrates the structure of ESH PDU. This PDU is constructed by an ES and may be broadcast to either ISs or another ES (if there is no IS identified) in the same subnetwork. It provides configuration information of the ES existence and reachability. The information carried in this PDU is the source NSAP (Network Service Access Point) address. NSAP address [1, 3] represents the point at which the network layer service can be accessed. Therefore, when there are multiple transport layer entities using the network layer service, the ESH PDU carries multiple addresses as well. The format of fixed and options part in this PDU apply to all other types of PDU. However, the fixed part of RA PDU has its own content.

b. ISH (Intermediate System Hello) PDU
This PDU is constructed by an IS and broadcast to ESs in the same subnetwork. Similar to the ESH PDU, it is used to provide configuration information about the IS existence and reachability. Consequently, it carries the address of the IS Network entity, which is also known as the IS NET [6]. Regarding this, there is only one address in this type of PDU.

c. RD (Redirect) PDU
This PDU is constructed by an IS and sent to ES when it discovers a better path for transmitting the CLNP PDU [6] originated by that ES. The better path could be the destination address of the CLNP PDU, which means the source ES is in fact directly connected to the destination ES through the same subnetwork. In this case, the ES-RD PDU carries the route redirection information consisting of the destination address of the CLNP PDU and the subnetwork address of the destination ES as the better next hop. Another redirection is to another IS which is also directly connected to the source ES. Therfore, the IS-RD PDU carries the destination address of the CLNP PDU, the subnetwork address and the NET of the IS to which the ES is redirected.

d. RA (Request Address) PDU
This PDU is constructed by an ES and broadcast to ISs in the same subnetwork when the ES does not have knowledge of its own address. Since this PDU is intended to send request only, it carries no address and does not need holding timer.

e. AA (Assign Address) PDU
This PDU is constructed by an IS and sent to ES upon a receipt of RA PDU. Through this PDU, the IS supplies the requesting ES with temporary NET, and it is the address carried by this PDU.

**3.2. Structure of Routing Information Base**

The information carried by the ES-IS Routing Information Exchange Protocol PDU is stored in a routing information base maintained in each ES and IS. To date, the structure is declared as simplest as possible to accommodate the storage of information from ESH and ISH PDU only.

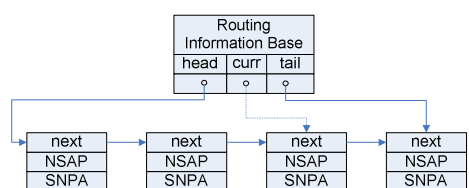

Figure 2. Routing Information Base

As shown in figure 2, the routing information base has the form of linked list. The pointer *head* holds the memory address of the first entry, while *tail* holds the address of the last entry. Another pointer, *curr*, is used to refer to any entry. An integer variable, *num_of_entry*, stores the total number of entries stored. Each entry has a pointer *next* that stores the memory address of the next entry, so that all entries can always be traced. It also stores the most basic configuration information, consisting of an NSAP address and an SNPA (Subnetwork Point of Attachment) [6].

# 4. Design of ES-IS Routing Information Exchange Protocol

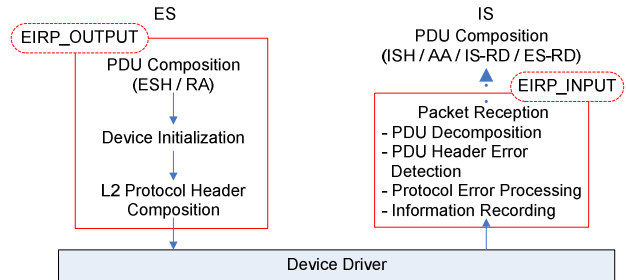

**Figure 3. Design of ES-IS Routing Information Exchange Protocol (ES to IS)**

Figure 3 shows the flow of PDU transmission from an ES to an IS. The details are as follows:
a. EIRP (ES-IS Routing Protocol) Output
   The output process begins with PDU composition. ESH PDU is composed periodically based on ES local configuration timer while RA PDU is composed when the ES needs to obtain its own NET. First, the fixed part is built by assigning values to its fields as regulated by ISO 9542 [6]. The composition of address part of ESH PDU header goes through a loop for multiple source NSAP addresses. There is no such process in RA PDU composition. The process may continue with options part construction. In ESH and RA PDU, the possible option parameters are security and priority. After all parts are constructed, the MSB and LSB of checksum value are generated by processing the header octets one by one. After composing the PDU, the protocol registers the Ethernet protocol ID in the same socket buffer structure used to send the packet and sets the content of data link layer protocol header consisting of physical source and destination addresses, and type. For transmission, the socket buffer structure is passed to device driver.

   The output process of ESH PDU is handled by report configuration function, and the one of RA PDU is handled by request address function. Report configuration function in ES is implemented in *eirp_build_and_send_esh*, which calls *eirp_build_fixed*, *eirp_build_esh_addr*, *eirp_buid_options*, *eirp_gen_csum*, *eirp_output*, and *eirp_finish_output* methods.

b. EIRP Input
   At the receiver IS, the received PDU is decomposed. First, the fixed part is extracted from the PDU by decomposing it into each single field. The protocol checks whether the NLPID is 130 and the version/protocol id extension is 1 [6]. If either value is wrong, the PDU is ignored. Through PDU header error detection function, the protocol ensures the consistency of the PDU content when it is sent and received. Any invalid modification in the header, which is certainly not followed by checksum adjustment, can be detected by checking the checksum parameter value using specific algorithm. The PDU header error detection is done early after receipt of PDU so that error PDU can soon be discarded. It is implemented in *eirp_check_csum* method.

   If no PDU header error is detected, the protocol examines header length, reserved octet and bits, and PDU type code in fixed part. Any invalid value among these fields is considered as protocol error, causing the PDU to be discarded. If the PDU passes this checking successfully, its address part is decomposed. In ESH PDU, it is checked whether the number of source address is not zero. Then, the conformance of each source address length and value to ATN addressing convention is also checked. If there is no error in address part, it continues with options part decomposition and examination. Each type of option has its own specification. Any invalid option parameter code, length, or value or duplicate parameter is considered as protocol error. All of these processes are executed in *eirp_rcv*, *eirp_rcv_finish*, and *eirp_rcv_options* methods.

Only the information from error-free PDU will be saved in the routing information base. The configuration information from ESH PDU is already extracted in address part decomposition. The IS records the {NSAP, SNPA} pair in local routing information base and replaces any information for identical pair through record configuration function. Any entry in the routing information base will be removed by certain flush function once the holding timer expires.

When an IS receives an ESH PDU and discovers that the sender ES is newly available, through configuration notification function, it sends an ISH PDU back to that ES. AA PDU is composed through assign address function if the IS receives RA PDU. In contrast, RD PDU is not constructed to response a PDU composed by the ES-IS Routing Information Exchange Protocol. Instead, it is composed if the IS receives a CLNP PDU from an ES and after certain examination figures out that the ES actually can bypass it. It is done by request redirect function.

From IS to ES, the packet processing can be described as follows:
a. EIRP Output
   The whole output process in IS is basically similar to the one in ES. The only difference is the types of PDU that both can construct. Like ESH PDU, ISH PDU is also composed periodically based on the IS local configuration timer, in addition to providing configuration information to newly available ES. Here, the report configuration function is implemented in *eirp_build_and_send_ish*, which calls *eirp_build_ish_addr* instead of *eirp_build_esh_addr*.

b. EIRP Input
   When a PDU is received by the ES, it passes through the procedures as in input process in IS. The configuration information provided by ISH PDU, which is a {NET, SNPA} pair, is stored by record configuration function. Both record configuration in ES and IS are implemented in *eirp_insert_entry* method. If the ES discovers that the sending IS is newly available, it sends an ESH PDU back to the IS. This is also the configuration notification function's role. When the received PDU is an AA PDU, the ES runs record address function to store its own NET. The receiver ES also stores the redirection information through record redirect function. When there are traffics with opposite direction through identical path, the redirection information's holding timer is refreshed by refresh redirect function so that it can still be used by either transmission.

   In addition to security and priority option parameters, RD PDU may also contain address mask and SNPA mask option parameters. Moreover, ISH PDU can suggest a configuration timer to ES through ES Configuration Timer (ESCT) option parameter [6]. Consequently, protocol error processing in ES must be complemented to check these types of option parameter.

## 5. Implementation

In this research, the protocol was implemented using C language for use in Linux Kernel. We choose Linux as the basis of our research because its kernel is modifiable and references are widely available. Version 2.6 was used because it is the latest compatible version for our research. In the existing Linux Kernel source code, a common structure called *sk_buff* (socket buffer) is used by data link layer, network layer, and transport layer to store the pointers to their protocol headers, respectively in *mac*, *nh*, and *h* union. It is declared in the *include/linux/skbuff.h* include file. Pointers to structure of ESH PDU header and ISH PDU header are also included in the *nh* union. The program of the protocol was inserted into Linux Kernel 2.6 by locating the folder of the source code files (*clnp_route*) in */usr/src/linux-2.6.xx/net*, where the folders of *ipv4* and *ipv6* modules are also stored. The header file, *clnp_route.h*, is located in */usr/src/linux-2.6.xx/include/linux*. We created a kernel module (*eirp_module*) to run the program inside the kernel. This kernel module requires initialization and termination functions. Then a *Makefile* was created to compile all source codes files. "Make" command is used to compile the program so that the *.ko* file is formed. We loaded the module temporarily in Linux kernel using "insmod" command. To terminate the module, we use "rmmod" command. The result of the program execution can be seen in */var/log/messages*.

For testing purpose, the following scenarios have been set up for each implemented function:
a. The report configuration function was tested by providing dummy address parameter and option parameters as the input for PDU composition function.
b. The PDU header error detection function was tested by intentionally modifying any random octet in the composed PDU header which checksum value has been generated. It was done to simulate malicious alteration of transmitted PDU.
c. The protocol error processing function was tested by assigning invalid values to certain fields of the PDU header before the generation of checksum parameter value.
d. The record configuration function was tested by creating dummy routing information base with two conditions. First, one of its entries was set to have identical address value as carried by the composed PDU. Second, all of the addresses in the routing information base and composed PDU were assigned with different values.

## 6. Conclusion

Regarding the protocol functions implementation, we conclude that the report configuration function has been successfully implemented to construct ESH and ISH PDU and include them into the socket buffer structure. Additionally, the developed PDU header error detection function can detect invalid modification in PDU header. Furthermore, the implemented protocol error processing function is able to detect invalid value in PDU composition. In addition, the record configuration function can store the address parameter carried by a PDU into the routing information base according to its newness. For future work, the module of the ES-IS Routing Information Exchange Protocol must be complemented with the functions that are not yet developed and integrated with the full CLNP modules. Additional test cases for performance measurement must also be executed.

**References**
[1] APANPIRG/ATNTTF Ad Hoc Group: "PROPOSED FINAL DRAFT OF ASIA/PACIFIC ATN ADDRESSING PLAN", International Civil Aviation Organization, Singapore, 2001.
[2] Dhas, C., Mulkerin, T., Wargo, C., Nielsen, R. & Gaughan, T.: "Aeronautical Related Applications Using ATN and TCP/IP Research Report", Computer Networks and Software, Inc., Virginia, 2000.
[3] Fans Information Services: "Comprehensive ATN Manual (CAMAL), Part 1 Introduction and Overview", FANS Information Services Ltd., 1999.
[4] ICAO: "ASIA/PACIFIC REGIONAL PLAN FOR THE NEW CNS/ATM SYSTEMS", International Civil Aviation Organization Asia and Pacific Office, Bangkok, 2005.
[5] ISO/IEC: "INTERNATIONAL STANDARD ISO/IEC 8473-1 Information Technology – Protocol for Providing the Connectionless-Mode Network Service", $2^{nd}$ edition, ISO/IEC Copyright Office, Geneve, 1998.
[6] ISO/IEC: "INTERNATIONAL STANDARD ISO/IEC 9542: Information technology — End system to intermediate system routeing information exchange protocol for use in conjunction with the protocol for providing the connectionless-mode network service (ISO/IEC 8473)", $2^{nd}$ edition, ISO/IEC Copyright Office, Geneve, 1994.
[7] NetBSD: "NetBSD CVS Repositories". Accessed on July 2007. Available at http://cvsweb.netbsd.org/bsdweb.cgi/
[8] Rio, M., Goutelle, M., Kelly, T., Hughes-Jones, R., Martin-Flatin, J. P. and Li, Y. T.: "A Map of the Networking Code in Linux Kernel 2.4.20", March 2004.
[9] Signore, T.L & Girard, M.: "THE AERONAUTICAL TELECOMMUNICATION NETWORK (ATN)", The MITRE Corporation, Bedford, 1998.
[10] The Agency for The Assesment and Application of Technology (BPPT): "Technology Roadmap of The Agency for BPPT", year 2007-2015, Jakarta, March 2007.